\documentclass[aps,twocolumn,floats,prl,nofootinbib,superscriptaddress,10pt]{revtex4-2}

\usepackage[utf8]{inputenc}

\usepackage{graphicx,amsmath,amsfonts,amssymb,slashed}
\usepackage{bbold,wasysym}
\usepackage{ulem}

\usepackage{amsmath,amsfonts,bbm,euscript,hyperref}

\usepackage{xcolor}
\usepackage{color}
\usepackage{hyperref}
\usepackage{bm}

\hypersetup{colorlinks, citecolor=blue, linkcolor=red, urlcolor=blue}

\usepackage{graphicx,capt-of}

\newcounter{subfloat}

\newcounter{subfloat2}

\newcommand{\nco}{\newcommand}
\nco{\beq}{\begin{equation}} \nco{\eeq}{\end{equation}}
\nco{\beqa}{\begin{eqnarray}} \nco{\eeqa}{\end{eqnarray}}
\def\be{\begin{equation}}
\def\ee{\end{equation}}    
\def\baray{\begin{eqnarray}}
\def\earay{\end{eqnarray}}

\nco{\lra}{\leftrightarrow}

\nco{\sss}{\scriptscriptstyle} \nco{\dphi}{\varphi}
\nco{\lsim}{\mbox{\raisebox{-.6ex}{~$\stackrel{<}{\sim}$~}}}
\nco{\gsim}{\mbox{\raisebox{-.6ex}{~$\stackrel{>}{\sim}$~}}}

\def\IK{\relax{\rm I\kern-.20em K}}
\def\IM{\relax{\rm I\kern-.20em M}}

\def\lsim{\mbox{\raisebox{-.6ex}{~$\stackrel{<}{\sim}$~}}}
\def\gsim{\mbox{\raisebox{-.6ex}{~$\stackrel{>}{\sim}$~}}}
\def\sss{\scriptscriptstyle}

\def\ga{\mathrel{\raise.3ex\hbox{$>$\kern-.75em\lower1ex\hbox{$\sim$}}}}
\def\la{\mathrel{\raise.3ex\hbox{$<$\kern-.75em\lower1ex\hbox{$\sim$}}}}

\def\Mpl{M_p}

\begin{document}

\begin{flushright}   CTPU-PTC-23-29  \end{flushright}

\title{
Axion-Gauge Dynamics During Inflation 
\\
as the Origin of Pulsar Timing Array Signals and Primordial Black Holes 
%T1:Inflationary Axion-Gauge Interactions Producing PBHs and GW at PTA Data
%\\
%T2:Axion Inflation Sourced Stochastic Gravitational Wave Evidence at Pulsar Timing Arrays and Primordial Black Holes
}
\author{Caner \"Unal}
\email{cerul2870@gmail.com}
\email{unalx005@umn.edu}
\affiliation{Department of Physics, Ben-Gurion University of the Negev, Be’er Sheva 84105, Israel}
%\affiliation{Department of Physics, Middle East Technical University, 06800 Ankara, Turkey}
\affiliation{Feza Gursey Institute, Bogazici University, Cengelkoy, Istanbul, Turkey
}

\author{Alexandros Papageorgiou}
\email{papageo@ibs.re.kr}
\email{a\_papageorgiou@yahoo.gr}
\affiliation{Particle Theory and Cosmology Group, Center for Theoretical Physics of the Universe, 
Institute for Basic Science (IBS), 34126 Daejeon, Korea}

\author{Ippei Obata}
\email{ippei.obata@ipmu.jp}
\affiliation{Kavli Institute for the Physics and Mathematics of the Universe (Kavli IPMU, WPI),UTIAS, The University of Tokyo, 5-1-5 Kashiwanoha, Kashiwa, Chiba, 277-8583, Japan}

\begin{abstract}We demonstrate that the recently announced signal for a stochastic gravitational wave background (SGWB) from pulsar timing array (PTA) observations, if attributed to new physics, is compatible with primordial GW production due to axion-gauge dynamics during inflation. More specifically we find that axion-$U(1)$ models may lead to sufficient particle production to explain the signal while simultaneously source some fraction of sub-solar mass primordial black holes (PBHs) as a signature. Moreover there is a parity violation in GW sector, hence the model suggests chiral GW search as a concrete target for future. We further analyze the axion-$SU(2)$ coupling signatures and find that 
in the low/mild backreaction regime, it is incapable of producing PTA evidence and the tensor-to-scalar ratio is low at the peak, hence it overproduces scalar perturbations and PBHs.

%We show that the recent pulsar timing array evidence for stochastic gravitational wave (GW) background, which might imply beyond astrophysical sources, can be explained with a primordial GW signal resulting from axion and gauge field interactions. Interestingly, due to the high efficiency of particle production in axion inflation, such models can evade primordial black hole (PBH) over-production bounds compared to scalar induced GW, and they can even source some fraction of solar and sub-solar mass PBHs as a signature. There are more definitive signatures of such interactions including violation of parity in GW, ie helical GW, hence the chirality of the detected background from PTAs could be a smoking gun of this primordial universe phenomena.
\end{abstract}

\maketitle

{\bf Introduction.} There is a strong evidence for a stochastic gravitational wave background (SGWB) in Pulsar Timing Array (PTA) data from the NANOGRav \cite{NANOGrav:2023gor,NANOGrav:2023hvm}, EPTA \cite{Antoniadis:2023ott,Antoniadis:2023zhi}, PPTA \cite{Reardon:2023gzh} and CPTA \cite{Xu:2023wog} experiments in the nHz frequency regime.
While this observed signal is mainly thought to be of standard astrophysical origin sourced by supermassive black hole binary mergers \cite{NANOGrav:2020spf,Middleton:2020asl,Broadhurst:2023tus}, %there is a \textcolor{red}{mild tension between NANOGrav data and such a model \cite{NANOGrav:2023hvm},
on top of astrophysical background, there may be a possibility that the data also implies SGWB of cosmological origin, such as cosmic strings \cite{Ellis:2020ena,Blasi:2020mfx,Samanta:2020cdk,Ellis:2023tsl,Wang:2023len}, cosmological phase transitions \cite{Higaki:2016jjh,Kobakhidze:2017mru,Arunasalam:2017ajm,Nakai:2020oit,Ratzinger:2020koh,Neronov:2020qrl,Chiang:2020aui,NANOGrav:2021flc,Ferreira:2022zzo,Ashoorioon:2022raz,Han:2023olf,Li:2023yaj,Fujikura:2023lkn,Kitajima:2023cek,Athron:2023mer,Blasi:2023sej,Lu:2023mcz,Li:2023bxy,Ghosh:2023aum}, scalar induced GW (SIGW) \cite{Chen:2019xse,Wang:2019kaf,Vaskonen:2020lbd,DeLuca:2020agl,Kohri:2020qqd,Domenech:2020ers,Zhao:2022kvz,Franciolini:2023pbf,Inomata:2023zup,Liu:2023ymk,Wang:2023ost,Cai:2023dls}.

How about a possibility of primordial gravitational waves (GW) from cosmic inflation \cite{guth:1981,sato:1981,linde:1982,albrecht/steinhardt:1982}?
%The standard inflationary scenario tells us that the resultant tensor power spectrum is nearly scale-invariant but slightly red-tilted.
In the standard inflationary model, the amplitude of SGWB is nearly scale-invariant and is parameterized by a ratio of tensor and scalar power spectra, called tensor-to-scalar ratio $r$.
It is yet to be detected, and current cosmic microwave background (CMB) experiments, Planck \cite{Tristram:2020wbi} and BICEP/Keck \cite{BICEP2:2018kqh}, have put an upper bound $r \lesssim 0.034$ at 95 \% C.L. \cite{Tristram:2021tvh}. This CMB bound predicts the SGWB much smaller than that measured in PTA data. Explaining the data requires a blue-tilted spectrum around the nHz frequency regime \cite{NANOGrav:2023hvm,Vagnozzi:2023lwo,Datta:2023vbs}, which is not compatible with the standard inflationary scenario producing nearly scale invariant SGWB.

In the simplest scenario discussed above, SGWB is produced by the vacuum fluctuation of the metric, and the tensor-to-scalar ratio is directly related to the energy scale of inflation. However, this is not necessarily true if additional matter fields could source the GW during inflation. In such scenarios, the dynamics of gauge fields coupled to axions, both Abelian ($U(1)$) \cite{sorbo:2011,Barnaby:2011vw,cook/sorbo:2012,Barnaby:2011qe,barnaby/etal:2012,cook/sorbo:2013,Mukohyama:2014gba,Ferreira:2014zia,Peloso:2015dsa,namba/etal:2015,Domcke:2016bkh,Garcia-Bellido_2016,Obata:2016oym,Ozsoy:2020ccy} and non-Abelian ($SU(2)$) \cite{Maleknejad:2011jw,Dimastrogiovanni:2012ew,Adshead:2013nka,Adshead:2013qp,obata/soda:2016,Obata:2016xcr,Dimastrogiovanni_fasiello_fujita_2016,Agrawal_etal,adshead/martinec/sfakianakis:2016,adshead/sfakianakis:2017,Thorne,agrawal/fujita/komatsu:2018b,domcke/etal:2019,fujita/sfakianakis/shiraishi:2019,Fujita:2021flu} groups, have been well studied (see \cite{Maleknejad_etal,Komatsu:2022nvu} for reviews).
Axions generically couple to gauge fields via topological Chern-Simons interactions, and this coupling violates the conformal invariance of the gauge field. Owing to this coupling, there is copious production of one circular polarization of the gauge field which in turn amplifies GW background during inflation. The resultant sourced SGWB is generically scale-dependent and its spectral shape is controlled by the time evolution of axion field. Therefore, depending on the potential, the sourced SGWB is enhanced on intermediate scales during inflation and is potentially testable with future interferometers or PTAs \cite{cook/sorbo:2012,Domcke:2016bkh,Garcia-Bellido_2016,Obata:2016oym,Ozsoy:2020ccy,obata/soda:2016,Thorne,Campeti:2020xwn}.
The amplified gauge field also inevitably sources scalar density modes, aside from GW, which could in turn lead to the generation of primordial black holes (PBHs) after inflation \cite{Garcia-Bellido_2016,Domcke:2017fix,Garcia-Bellido:2017aan,Ozsoy:2020ccy,Ozsoy:2020kat} (see also PBHs via axions \cite{Guo:2023hyp}).

In this {\it Letter}, we show the possibility that the current PTA data can be explained by the primordial GW sourced by axion-gauge dynamics during inflation.
We consider two scenarios: $U(1)$ and $SU(2)$ models. The axion is a spectator field and realizes localized gauge field amplifications on intermediate scales. We discuss the possibility of generating a enhanced power spectrum of tensor modes compatible with the NANOGrav data from these models while satisfying some theoretical consistencies. We limit our analysis to the low/mild backreaction regime for which we have robust analytic expressions and remain agnostic about the dynamics in the strong backreaction regime, which require approximate numerical solutions \cite{Cheng:2015oqa,Notari:2016npn,Sobol:2019xls,DallAgata:2019yrr,Domcke:2020zez,Gorbar:2021rlt} or full simulations on the lattice \cite{Caravano:2022epk,Figueroa:2023oxc}. We further discuss a possibility that the sourced power spectrum of scalar modes can lead to scalar induced GW and PBH generation \cite{Garcia-Bellido:2017aan,Unal:2020mts,Unal:2018yaa}.
\\
\\
{\bf PTA Data.} %{\bf Can you please digitize NANOGRAV+EPTA data with error bars, they are the best results until now, we need a figure something like this https://arxiv.org/pdf/2306.17136.pdf}
PTA experiments found evidence for the characteristic strain amplitude as \cite{NANOGrav:2023gor,NANOGrav:2023hvm,Antoniadis:2023zhi})
\begin{equation}
h_c(f) = 10^{-14.3\pm0.3} \left( \frac{f}{yr^{-1}} \right)^{\gamma}, %\quad (f \sim \rm nHz) \ ,
\end{equation}
where $-0.5< \gamma<0.5$ in 2-$\sigma$ error bars.
In terms of the GW energy density, we have
\begin{equation}
\Omega_{GW,0} h^2 \simeq \left( \frac{g_0}{g_{*}} \right)^{1/3} \frac{2 \pi^2}{3 H_0^2} h_c^2 f^2 \ ,
\end{equation}
where $g_{*}$ and $g_0$ are the relativistic degrees of freedom at the time of GW formation and at present, $H_0$ is current Hubble parameter. Then, we evaluate $\Omega_{GW,0}\sim 10^{-7}$ which implies for the tensor power spectrum that $P_h\sim 10^{-2}$ around frequency $yr^{-1}$, and $\Omega_{GW,0}\sim 10^{-8}$ implying $P_h\sim 10^{-3} $ for $f\sim(3 \,  yr)^{-1}$ \footnote{ although PTAs have very-low sensitivity beyond $f > (2-3 yr)^{-1}$}. This is many orders of magnitude larger than current bounds on CMB scales, ie. $P_{h} (k_{CMB}) < 10^{-11}$, that is tensor power spectrum at CMB scales. Hence the background needs to be amplified at least 9 orders of magnitude compared to fluctuations at large scales.
\\
\\
{\bf Axion and U(1) Coupling.} We present an inflationary model where an axion field couples to a $U(1)$ gauge field with field strength tensor $F_{\mu\nu} = \partial_\mu A_\nu - \partial_\nu A_\mu$.
In this work, we consider the model where the axion coupled to the gauge field is a spectator \cite{Mukohyama:2014gba,namba/etal:2015}.
The Lagrangian density is as follows:
\begin{align}
\mathcal{L} &= \mathcal{L}_{\rm GR} + \mathcal{L}_{\rm inflaton} + \mathcal{L}_{\rm gauge} + \mathcal{L}_{\rm spectator}  \label{eq: Lag} \ , \\
\mathcal{L}_{\rm spectator} &= - \dfrac{1}{2}\partial_\mu\chi\partial^\mu\chi - U(\chi) - \lambda\dfrac{\chi}{4 f_a}F_{\mu\nu}\tilde{F}^{\mu\nu} \label{eq: Lags} \ ,
\end{align}
where $\mathcal{L}_{\rm GR} = \Mpl^2R/2$ and $\mathcal{L}_{\rm inflaton} = -(\partial\phi)^2/2 - V(\phi)$ represent the Lagrangian densities of Einstein-Hilbert action and a canonical inflaton action.
Regarding the inflaton potential, we let it unspecified and do not solve the background evolution of inflaton.
The lagrangian density of gauge field is defined as $\mathcal{L}_{\rm gauge} = -F_{\mu\nu}F^{\mu\nu}/4$.
%We approximate that the Hubble parameter $H$ is constant.
The Hodge dual of field strength is defined as $\tilde{F}^{\mu\nu} \equiv\sqrt{-g}\epsilon^{\mu\nu\rho\sigma}F_{\rho\sigma}/2$, where $\epsilon^{\mu\nu\rho\sigma}$ is an antisymmetric tensor satisfying $\epsilon^{0123} = g^{-1}$. The important thing is that the inflaton is not directly coupled to the gauge field \footnote{Axion being inflaton with a polynomial potential cannot generate PTA signal due to bounds on the primordial non-Gaussianity at CMB scales \cite{Niu:2023bsr}, CMB scales requires slower axion and this prevents enough enhancement at PTA scales, a way out by chaning slope around PTA scales is discussed in \cite{Garcia-Bellido_2016}.}. This assumption enables the generation of curvature perturbations sourced by the gauge field to be suppressed.

We consider the time evolution of the gauge field coupled to a rolling axion. To do it, we decompose the gauge potential $A_i$ into operators with two circular polarization modes in Fourier space $\hat{A}^\pm_{\bm{k}}$.
%\begin{align}
%A_i(t, \bm{x}) &= \sum_{\lambda = \pm}\int\dfrac{d\bm{k}}{(2\pi)^3}\hat{A}^\lambda_{\bm{k}}e^\lambda_i(\hat{\bm{k}})e^{i\bm{k}\cdot\bm{x}} \ , \\
%\hat{A}^\lambda_{\bm{k}} &= A^\lambda_k \hat{a}^\lambda_{\bm{k}} + A^{\lambda*}_k\hat{a}^{\lambda\dagger}_{-\bm{k}} \ , \qquad \left[\hat{a}^{\lambda_1}_{\bm{k}}, \ \hat{a}^{\lambda_2\dagger}_{-\bm{k}'} \right] = (2\pi)^3\delta^{\lambda_1\lambda_2}\delta(\bm{k} + \bm{k}') \ .
%\end{align}
Then, in terms of a dimensionless time variable $x \equiv -k\tau$, the equation of motion (EoM) for the mode function $A^{\pm}_k$ is given by
\begin{equation}
\partial_x^2 A^\pm_k + \left(1 \mp \dfrac{2\xi}{x}\right)A^\pm_k = 0 \ , \quad \xi \equiv \dfrac{\lambda\dot{\bar{\chi}}}{2f_a H} \ , \label{eq: AU1}
\end{equation}
where the dispersion relation is modified by the axion-$U(1)$ coupling controlled by a model parameter $\xi$.
%For a while, we neglect a time dependence of $\xi$ and assume that $\xi$ is constant.
Initially, when the size of the mode function is deep inside the horizon ($x \gg 1$), this correction term is negligible and the gauge field obeys the standard dispersion relation. 
When it becomes comparable to the horizon size, however, one circular polarization mode gets an effective negative mass square for $x \lesssim 2\xi$ and a growing solution appears.
The plus mode $A^+_k$ is amplified exponentially $A^+_k \propto e^{\pi\xi}$
%\begin{equation}
%A^+_k \simeq \dfrac{1}{\sqrt{2k}}\left(\dfrac{x}{2\xi}\right)^{1/4}e^{\pi\xi - 2\sqrt{2\xi x}}
%\end{equation}
in the time interval $(8\xi)^{-1} \lesssim x \lesssim 2\xi$.
After $x \lesssim (8\xi)^{-1}$, the amplification weakens and the energy density of gauge field is diluted away by the expansion of the universe \footnote{We note that $x \lesssim (8\xi)^{-1}$ is chosen as a typical scale for the integration limit where the production of gauge fields are stronger than dilution. This scale can be chosen differently and results
will be the same.} Therefore, the gauge field production takes place at around horizon-crossing and enhances other coupled fluctuations.

This amplified gauge field enhances the coupled metric tensor modes $g_{ij} = a^2(\delta_{ij} + h_{ij})$ at second order level via the transverse-traceless components of energy-momentum tensor of electromagnetic field.
%The EoM for $h_{ij}$ is given by
%\begin{equation}
%\dfrac{1}{2a^2}\left(\partial_\tau^2 + 2\mathcal{H}\partial_\tau - \partial_{\bm{x}}^2 \right)h_{ij} = %\dfrac{1}{\Mpl^2}\left(-E_iE_j - B_iB_j\right)^{TT} \ ,
%\end{equation}
%where $TT$ denotes the transverse and traceless projection of the energy-momentum tensor of electromagnetic field.
We define the power spectrum of GW
\begin{equation}
\langle h_{\bm{k}}h_{\bm{k}'} \rangle = (2\pi)^3\delta(\bm{k} + \bm{k}')\dfrac{2\pi^2}{k^3}\left(P^R_h + P^L_h \right) \ ,
\end{equation}
where $P^{R(L)}_h$ is a dimensionless right- (left-) handed tensor power spectrum.
%Without the backreaction effect of gauge field, we finally obtain 
%\begin{equation}
%P^{R/L}_h \simeq P^{(v)}_h\left(1 + P^{(v)}_hf_{R/L}(\xi)\exp(4\pi\xi)\right) \ .
%\end{equation}
%The second term is a sourced term of gauge field production.
%For $2 \lesssim \xi \lesssim 3$, the numerical factor $f_{R/L}(\xi)$ is approximated as \cite{Barnaby:2011vw}
%\begin{equation}
%f_R(\xi) \simeq \dfrac{4.3\times10^{-7}}{\xi^6} \ , \quad f_L(\xi) \simeq \dfrac{9.2\times10^{-10}}{\xi^6} \ .
%\end{equation}
%\begin{equation}
%f_R(\xi) \simeq \dfrac{2.6\times10^{-7}}{\xi^{5.7}} \ , \quad f_L(\xi) \simeq 10^{-3}f_R(\xi) \ .
%\end{equation}
%We notice that both modes are amplified.
The shape of power spectrum is related to the time evolution of model parameter $\xi(t_k)$, which is determined by the details of the axion potential.
In this paper, we follow the previous works \cite{namba/etal:2015} and adopt a cosine potential
\begin{equation}
U(\chi) = \Lambda^4\left[\cos\left(\dfrac{\chi}{f_a}\right) + 1\right] \ .
\end{equation}
%
%and consider the parameter space where the backreaction of gauge field is negligible.
The velocity of the field gets a maximum value around $\chi(t=t_*) = \pi f_a/2$, where $t_*$ is the moment of maximum velocity. In the case of axion-$U(1)$ model, the slow-roll solution is given as
\begin{align}
\xi = \dfrac{\lambda\delta}{2\cosh{[\delta H (t-t_*)]}} \ , \quad \delta \equiv \dfrac{\Lambda^4}{3H^2f_a^2} \ .
\end{align}
%
%Around the peak ($t \simeq t_*$), it is approximated by
%\begin{equation}
%\xi \simeq \xi_*\left(1-\dfrac{(\delta H(t-t_*))^2}{2}\right) \ , \quad \xi_* \equiv \dfrac{\lambda\delta}{2} \qquad (\text{axion-} U(1)) \ .
%\end{equation}

\begin{figure}
\hspace{-0.5cm}
\includegraphics[width=0.499\textwidth,angle=0]{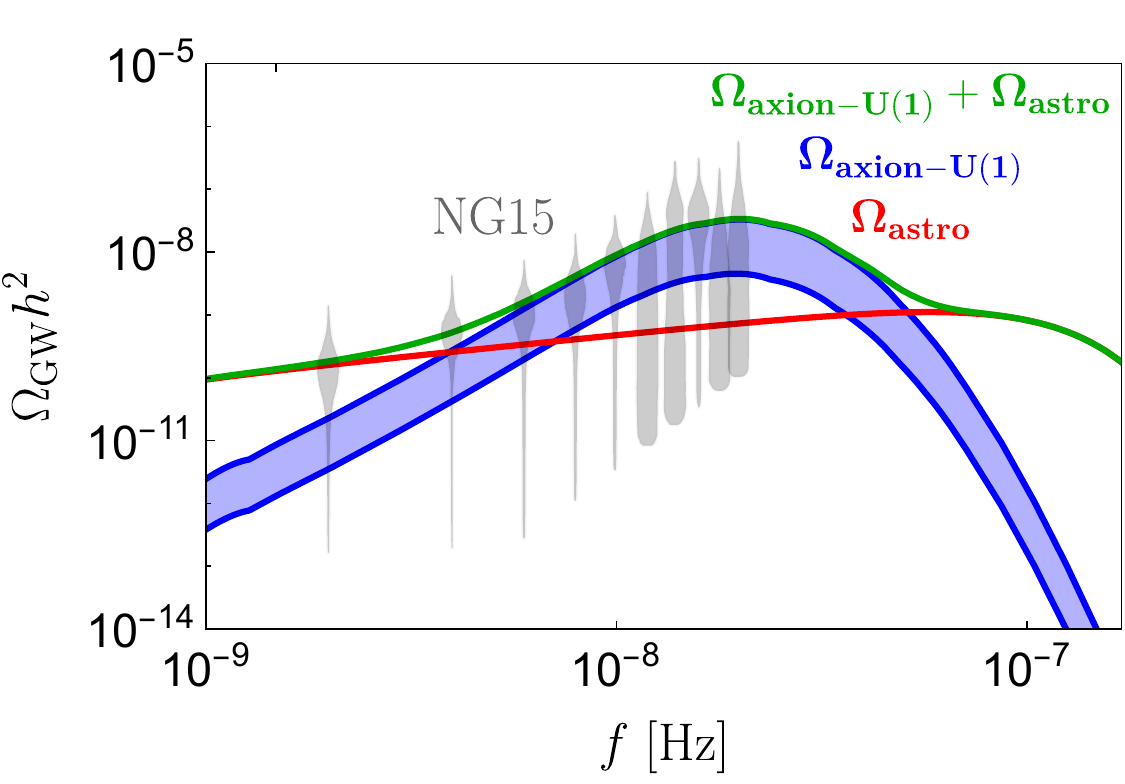}
\caption{
An example of GW production for the spectator axion-U(1) model that is in agreement with the NANOGrav signal. The chosen parameters are $\delta=0.5$, $6< \xi_*< 6.2$, and $r_v = 10^{-2}$ (blue). The astrophysical origin SMBHB (red), and the total SGWB (green) are also displayed.}
\label{figomegagw}
\vspace{-0.15in}
\end{figure}

{\bf GW Production for U(1).}
%{\it GW in U(1)}.
%In U(1) no linear coupling between tensor modes and gauge field which is true spin-1 field. Hence sourcing is done through 3-pt interaction h-A-A, and the tachyonically enhanced gauge fields also source axion(and indirectly inflaton even if inflaton is not axion via $\phi-A-A$)
In Figure \ref{figomegagw}, we plot the SGWB from axion-$U(1)$ model and compare it with that from astrophysical origin \cite{NANOGrav:2023gor,NANOGrav:2023hvm}. 
The astrophysical GWB from SMBHB is expected to have $10^{-15}$ characteristic strain amplitude, and -2/3 frequency slope, which translates into $\Omega_{\rm astro}\sim 10^{-9} \left(f/yr^{-1}\right)^{2/3}$. 
Additionally, these axion-gauge field interactions produce a unique signature, namely at small frequencies it scales as $f^3$ consistent with the NANOGrav slope, the peak is in log-normal shape and there is a rapid decay in the UV frequencies.

Power spectra, ie. two-point function, of sourced tensor and scalar modes, $P_{i,s}$, around their peak are parametrized as \cite{namba/etal:2015}
\begin{equation}
    P_{i,s} (k) = \left( \frac{H^2}{8 \pi^2 M_p^2}\right)^2 f_{2,i} \qquad (i = h, \zeta) \ ,
\end{equation}
where $f_{2,i}=A_i {\rm Exp}\left[-\ln\left(\frac{k}{x_ik_*}\right)^2/(2\sigma_i^2)\right]$ with the amplitude $A_i$, the spectral width $\sigma_i$, and with the spectral position $x_i$ at which the function has the bump $k = x_ik_*$.
Specifically, we have 
%$ A_h = {\rm exp} \[−6.85 + 9.05 \chi_* + 0.0596 \chi_*^2 \]$
$A_h= {\rm Exp}[-6.85 + 9.05 \xi_* + 0.0596 \xi^2_*]$
where $\xi_* = \lambda\delta/2$ is the largest value of $\xi(t=t_*)$ at the fastest motion point.
%Similarly for curvature perturbations
%\begin{equation}
%    P_{\zeta,s} (k) = \left( \frac{H^2}{8 \pi^2 M_p^2}\right)^2 f_{2,\zeta}
%\end{equation}
%where $f_{2,\zeta}=A_\zeta = Exp[−6.47 + 9.04\chi_* + 0.0586 \chi^2]$. 
We show in the shaded region the SGWB for particle production parameter in the range, $\xi_*=6.2$ to  $\xi_*=6$. We further confirm that primordial GW peak is greater than astrophysical for $\xi_*>5.8$.
%For $\xi_* = 6.1$ and $r_v = 10^{-2}$, where $r_v \equiv 16\epsilon_H$ is a tensor-to-scalar ratio from vacuum fluctuations, it leads to 
For the chosen example parameters, tensor-to-scalar ratio at the peak $r(k_p)\sim {\rm Exp}[-0.4]\sim 1/2$. For different rolling times and particle production parameters, see \cite{namba/etal:2015}.  

A value of $P_h\sim 10^{-3}$ \footnote{We note that due to the decay of vector modes during inflation, $N_{eff}$ bounds are only relevant for the GW background, not for the gauge field that forms $0.5-5\%$ of total energy density at its peak, which still satsifies $N_{eff}$ bounds with tiny margin.} corresponds to $P_\zeta\sim 10^{-2.5}$ which may not violate PBH bound if fluctuations are relatively Gaussian as we will discuss in the next section. Note that although the scalar perturbations are characterized by a narrowly peaked amplitude, the GW spectrum signal is scaled as $f^3$ in the low frequency regime, namely in $f\ll f_{\rm peak}$.

By imposing backreaction and perturbativity bounds from the semi-analytic calculation, for $\delta=0.5$, we obtain $\xi_*$ = 6.11 is the maximum value that can satisfy $f_a < M_p$ by using the results in \cite{Peloso:2016gqs,Campeti:2022acx}, that is, for smaller $\xi_*$, the decay constant of the axion can be chosen smaller than $M_p$. $\xi_*$ larger than this value requires the decay constant of the axion to be larger than $M_p$, which is not viable. For this parameter region, non-linear analysis becomes necessary and it is beyond the scope of our paper.
\\
\\
{\bf Axion and SU(2) Coupling.} Next, we present an inflationary model where an axion couples to an $SU(2)$ gauge field with field strength $F^a_{\mu\nu} = \partial_\mu A_\nu - \partial_\nu A_\mu -g\epsilon^{abc}A^b_\mu A^c_\nu$.
The original model is called chromo-natural inflation, where an inflaton is directly coupled to the $SU(2)$ field \cite{adshead/wyman:2012}. However, this model was found to be inconsistent with CMB observations \cite{Dimastrogiovanni_etal,Adshead_etal,adshead/etal:2013}\footnote{see also \cite{Bagherian:2022mau} for some challenges concerning UV completing the axion-$SU(2)$ model.}. In this paper, we adopt the spectator axion-SU(2) model \cite{Dimastrogiovanni_fasiello_fujita_2016}, where the Lagrangian density is given by the same form as the $U(1)$ model \eqref{eq: Lag}, \eqref{eq: Lags}.

At the background level, the $SU(2)$ field can acquire an isotropic background value \cite{maleknejad/sheikh-jabbari:2011,maleknejad/sheikh-jabbari:2013}
$\bar{A}^a_i(t) = a(t)Q(t)\delta^a_i$, which is diagonal between the indices of $SU(2)$ and $SO(3)$ algebra. Since this configuration respects the spatial isotropy, the background spacetime can be described by a simple FLRW metric. The axion-$SU(2)$ coupling in the presence of a non-zero isotropic gauge field vacuum expectation value, {\it vev}, induces a friction in the equation of motion for the axion. The {\it vev} is assumed to be at the bottom of its effective potential $Q(t) \simeq (-f_a U_\chi/(3g\lambda H))^{1/3}$ and the particle production is characterized by the parameter 
\begin{equation}
    m_Q = g Q/H
\end{equation}
related to the $U(1)$ parameter through $\xi \simeq m_Q + m_Q^{-1}$. This solution is an attractor even if we start from the initial anisotropic parameter space as long as it leads to a stable inflationary period \cite{maleknejad/erfani:2014,wolfson/maleknejad/komatsu:2020,Wolfson:2021fya}.

At the level of perturbations, the fluctuation of the gauge field $\delta A^a_i$ possesses the transverse-traceless mode $\delta A^a_i \supset t^a_i$ owing to the background rotational symmetry. In the presence of the background vev, the tensor modes of the gauge field couple to GW at the linear level. In an analogous manner to the $U(1)$ case only one chirality of the tensor perturbations experiences a tachyonic instability around horizon-crossing. Hence, due to the linear coupling, only one chirality mode of GW is amplified. In this paper, we assume the amplified mode is right-handed: $t_R$ and $h_R$. The ratio of the sourced fluctuations, $P_{h,s}$, to the vacuum fluctuations, $P_{h,v}$, takes the form \cite{Dimastrogiovanni_fasiello_fujita_2016}
\begin{equation}
R_{\rm GW}\equiv\frac{P_{h,s}}{P_{h,v}}=\frac{\epsilon_B}{2}{\cal F}^2
\label{eqsugw}
\end{equation}
where $\epsilon_B \equiv g^2Q^4/(HM_p)^2=(m_Q^4/g^2) \; (H^2/M_p^2)$ and $\mathcal{F}$ is a numerical factor and in the parameter range $m_Q \gtrsim 2$ it is approximated by 
$\mathcal{F}^2 \simeq e^{3.6m_Q}$ \cite{Dimastrogiovanni_fasiello_fujita_2016}. Combining all, we express the primordial SGWB as follows
\begin{equation}
    \Omega_{GW,0}h^2= \frac{\Omega_{\gamma,0}h^2}{24}   \left( \frac{g_0}{g_{*}} \right)^{1/3} P_{h,s}
\end{equation}
where $\Omega_{\gamma,0}$ is radiation energy density, and $g_0$ and $g_*$ are the relativistic degrees of freedom today and re-entry.

\begin{figure}
\hspace{-0.5cm}
\includegraphics[width=0.499\textwidth,angle=0]{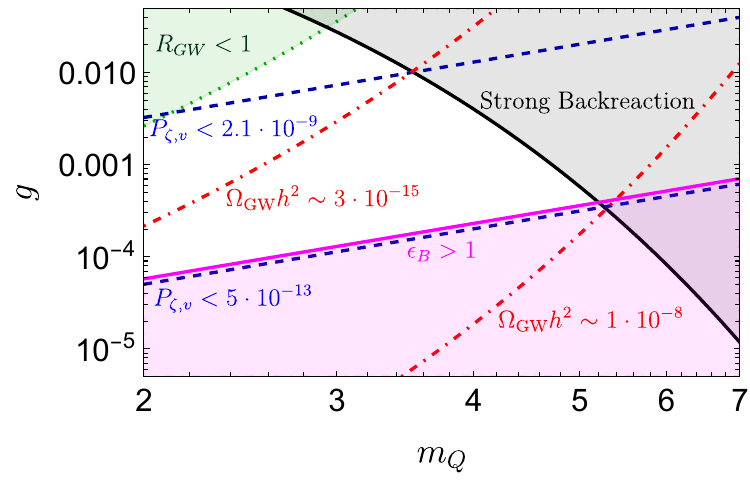}
    \caption{Constraints for the case of the axion-SU(2) model. The black contour displays the regime for which the backreaction is relevant (\ref{eq:su2backreaction}) and therefore beyond the scope of this work. The two blue dashed lines indicate the corresponding value of the vacuum contribution to the power spectrum of scalar perturbations (\ref{eq:su2pzmax}). The two red lines display the level of sourced GW production assuming $P_{h,v}$ the maximum allowed from CMB observations. Finally the magenta contour is disallowed because it would require for the slow-roll parameter to be greater than unity.}
    \label{fig:su2-constraints}
\end{figure}

For the non-Abelian case in the weak backreaction regime, it can be shown that the sourced GW can not be compatible with the NANOGrav signal. The main challenge of achieving the desired signal is due to the severe restriction imposed by the backreaction constraint. Unlike the axion-$U(1)$ case, the backreaction generally affects the EoM of the gauge field {\it vev} first (which is not present in the axion-$U(1)$ case) before there is a chance to backreact in the EoM of the axion, hence there are tighter constraints on the particle production parameter $m_Q$ compared to the Abelian equivalent $\xi$ \cite{maleknejad/komatsu:2019,Peloso:2016gqs}. We extend the analysis given in detail in section 5.2 of \cite{papageorgiou/peloso/unal:2019}. We present our combined constraints in Figure \ref{fig:su2-constraints} which is valid for small scales and contains the SGWB amplitude. We show two SGWB background values with red dashed-dotted lines, one is $\Omega_{\rm GW}  h^2\sim 3\cdot 10^{-15}$ that is the amplitude one gets in the weak backreaction regime and the scalar vacuum perturbations are taken at the face value of the CMB scales, the other one is $\Omega_{\rm GW}  h^2\sim 10^{-8}$, a GW signal of the order of the NANOGrav signal.

We impose the same backreaction constraint 
\begin{eqnarray}
    g\ll \left(\frac{24\pi^2}{2.3\cdot {\rm e}^{3.9 m_Q}}\frac{1}{1+\frac{1}{m_Q^2}}\right)^{1/2}
    \label{eq:su2backreaction}
\end{eqnarray}
which arises by demanding that the backreaction term in the EoM of the {\it vev} is smaller than the smallest background term, as in Ref. \cite{papageorgiou/peloso/unal:2019}. This is the safer condition due to a cancellation among the various terms as outlined in \cite{maleknejad/komatsu:2019} (see also \cite{Lozanov:2018kpk,Mirzagholi:2019jeb} for the Schwinger effect).

We could relax the normalization of $P_{\zeta,v}$ at small scales which is not constrained. The vacuum contribution to the power spectrum takes the form \cite{papageorgiou/peloso/unal:2019}
\begin{equation}
    P_{\zeta,v}\simeq \frac{H^2}{8\pi^2 M_p^2} \frac{\epsilon_\phi}{\sum_i \epsilon_i^2}=  \frac{g^2}{8\pi^2 m_Q^4}\frac{\epsilon_\phi \epsilon_B}{\left(\epsilon_\phi + \epsilon_B\right)^2}
\end{equation}
which further implies an absolute maximum value or $P_{\zeta,v}$ for a given $g$ and $m_Q$, namely
\begin{equation}
    P_{\zeta,v,{\rm max}}\simeq \frac{g^2}{32\pi^2 m_Q^4}
    \label{eq:su2pzmax}
\end{equation}
We plot the maximum attainable value of $P_{\zeta,v}$ for particular choices of $m_Q$ and g on Figure \ref{fig:su2-constraints} for reference. Naively one might think that values of $P_{\zeta,v}$ much lower than the CMB normalization are difficult to achieve, however that is not as clear cut in the non-Abelian model since there are several branches of solutions depending on the hierarchy of the slow roll parameters (for more details see Appendix F of \cite{papageorgiou/peloso/unal:2019}). In that case there is a lot more freedom to choose the various parameters, however for small $g$ and large $m_Q$ one inevitably enters the regime for which the slow-roll parameter $\epsilon_B$ is greater than one. Such a value is unacceptable as it would be incompatible with inflation\footnote{The "global" slow-roll parameter is the sum of the individual slow roll parameters $\epsilon_H=\epsilon_\phi+\epsilon_\chi+\epsilon_B+\dots$. Since they are all positive definite we can rule out the parameter space by requiring that any of them is greater than one.}. We plot some sample values of $P_{\zeta,v}$ in Figure \ref{fig:su2-constraints} and superimpose the corresponding sourced GW power spectrum for some sample choices. The upper red line is the maximum produced sourced GW allowed if we assume a flat spectrum for $P_{\zeta,v}$ from CMB to PTA scales and the lower red line is the parameter space that would account for the NANOGrav signal while being agnostic about the evolution of $P_{\zeta,v}$ at small scales. Note that we assumed for $P_{h,v}$ the maximum allowed value by CMB observations and a flat spectrum. This yields the absolute best case scenario and relaxing it makes the incompatibility even worse.

%{\bf Specifically, figure 3 of the same reference displays a triangle of acceptable values for the choice of parameters of the gauge coupling $g$ and particle production $m_Q$. }

Finally, we would like to point out that one can obtain an even more severe restriction in the parameter space displayed in Figure \ref{fig:su2-constraints} by investigating whether for the NANOGrav signal, at the peak the total amount of tensor-to-scalar ratio $r_{\rm tot}=\frac{P_{h,s}+P_{h,v}}{P_{\zeta,s}+P_{\zeta,v}}\sim \frac{P_{h,s}}{P_{\zeta,s}}\sim 1$ holds, which is a necessary condition for the non-overproduction of PBH. Specifically, using formulas (5.1) and (5.9) of \cite{papageorgiou/peloso/unal:2019} in which the sourced contributions dominate over the vacuum at the peak of the signal in the $\epsilon_B > \epsilon_\phi$ branch (Appendix F of the same reference), we have
 \begin{equation}
    r(k_{p}) \simeq \frac{8.5\cdot 10^{-2}}{N_k^2}\frac{\epsilon_B}{\epsilon_\phi} \;   \left( \frac{m_Q}{3.4}\right)^{-11} \; e^{-3.4 (m_Q-3.4)}\;.
\end{equation}
This ratio is always much smaller than one for $m_Q > 3.4$ which implies that the large values of $m_Q$ required to explain the NANOGrav signal are certain to overproduce scalar density perturbations, and expectedly PBH.

In summary, our analysis indicates that the NANOGrav signal observed is highly unlikely to be due to axion-$SU(2)$ dynamics during inflation in the low backreaction regime. It would be interesting to expand our analysis to the strong backreaction regime as in \cite{Ishiwata:2021yne}. However, such numerical analysis is beyond the scope of the current work.

{\bf PBH formation and Induced GW.} PBHs are produced in large abundances with higher efficiency, especially in the case of large non-Gaussianity. For same PBH abundance, the following relation holds between Gaussian and  chi-square ($\chi^2$) non-Gaussian density perturbations \cite{Garcia-Bellido:2017aan,Domcke:2017fix}
\begin{equation}
P_{\zeta,{\chi^2}} \sim \frac{2}{\zeta_c^2} \; P_{\zeta,G}^2 \ .
\end{equation}
$\zeta_c \sim \frac{9}{4}\delta_c$ is the critical threshold for curvature perturbations. Hence, for considerable PBH abundances we have $P_{\zeta,G}\sim 2\cdot 10^{-2}$, and  $P_{\zeta,\chi^2}\sim  10^{-3}$. For given amplitude and peak frequency, we estimate a PBH population, ranging from $10^{-10}-10^{-2}$ fraction of dark matter concentrated in masses about $10^{-4} - 10^{-1}M_\odot$
\footnote{Non-Gaussianity also affect the collapse threshold and perturbation shape, which are subdominant compared to the probability distribution function, hence neglected in this work.}.

It has been shown that as the fluctuations grow, they are highly non-Gaussian and approach the $\chi^2$ distribution. However, simulations \cite{Caravano:2022epk} conducted in the case of the axion-$U(1)$ model reveals that near the peak scalar perturbation statistics deviate from a $\chi^2$ distribution and converge to a Gaussian due to non-coherent addition of modes, thus weakening the PBH bound, allowing to some extent higher amplitude perturbations (See also recent results \cite{Figueroa:2023oxc} discovering the UV regime of momentum distribution.).  We expect the same trend to exist in the case of axion-$SU(2)$.

Enhanced scalar perturbations also source GW, scalar induced GW. In axion inflation models, the primordial GW spectrum is larger than the scalar induced GW background, which is also the case in our result, for given parameters $\Omega_{{\rm SIGW,\,peak}}\sim 10^{-9}$ as in \cite{Unal:2018yaa,Unal:2020mts,Garcia-Bellido:2017aan}, and we leave the spectrum details to future work.
 
%Then nonlinear relationship between density and curvature factor of 2

%This regime is particularly relevant for the recent NANOGrav result.
%Consequently, a more precise nonlinear analysis is required to obtain an accurate estimate for the amplitude of gravitational waves (GW) from the axion-U(1) model in the relevant observational regime. 

%Regarding the statistics becoming Gaussian has been proven for axion-U(1), but we expect it to be extended also to the SU(2) case because of the simple central limit argument.

{\bf Discussion and Conclusions.} The newly released PTA measurements show evidence of a SGWB. Although compatible with a background arising from supermassive black bole binary mergers, it is interesting to interpret the signal as having an early universe origin. 

%{\bf from abstract}
%We show that recent pulsar timing array evidence for stochastic gravitational wave(GW) background, which might imply beyond astrophysical sources, can be explained with primordial GW signal resulting from axion and gauge interactions. Interestingly, due to high efficiency of particle production in axion inflation, such models can evade primordial black hole (PBH) over-production bounds compared to scalar induced GW, and they can even source some fraction of solar and sub-solar mass PBHs as a signature. There are more certain signatures of such interactions including violation of parity in gravitational waves, ie helical GW, hence the helicity of the detected background from PTAs could be a smoking gun of this primordial universe phenomena.

We show that it is possible with axion-gauge field interactions during inflation. There are two main contributions to SGWB, one from sourced primordial GW background and the other from scalar induced GW (SIGW) background resulting from enhanced scalar perturbations. However, explaining the PTA signal with SIGW is a hard task due to PBH overproduction. Remarkably, in axion inflation, primordial GW production usually dominates over SIGW, namely primordial production is more dominant compared to SIGW, hence it allows a chance to explain the PTA data, and at the same time generating interesting signatures such sub-dominant fraction dark matter in primordial black holes and another SGWB, called scalar induced GW background.

We employ two models to explain the signal via this mechanism. The first one is an axion coupled to an Abelian gauge field \cite{namba/etal:2015,Garcia-Bellido_2016}. The axion rolls down for a finite period during inflation and when its speed is maximum, it sources one chirality of the gauge field. As a result the amplified gauge field sources one chirality of GW, axion and inflaton perturbations. The low frequency regime of the GW signal scales with $f^3$ instead of a log-normal fall, which improves the fit considerably together with the astrophysical background.

The second model is an axion coupled to a non-Abelian gauge field in which  due to the non-zero, isotropic, vacuum expectation value of the gauge field, there is a linear coupling between GW and the gauge field \cite{adshead/etal:2013,Dimastrogiovanni_fasiello_fujita_2016}, and the scalar modes are sourced via a cubic coupling \cite{papageorgiou/peloso/unal:2018,papageorgiou/peloso/unal:2019}. The model requires  large amplification of gauge modes such that GW can explain the PTA data, but this amplification results in two potential pitfalls i) strong backreaction  and ii) low tensor-to-scalar ratio at the peak, both of which are very difficult to overcome.
%
%\begin{enumerate}
%\item Such large fluctuations can only be enhanced by a large particle production parameter, which in almost all cases sets us in the strong backreaction regime which signals the breakdown of our approximate solutions. Our analysis indicates that there is no possible way to realize the PTA measured SGWB amplitude in the low backreaction regime. We are careful not to extend that conclusion in the strong backreaction regime, however a naive guess would be that in the strong backreaction regime the tensor mode instability is quenched and therefore it would be difficult to produce such a large signal, however it is possible that the mechanics of the strong backreaction are highly non-trivial and hard to predict.
%
%\item Even though scalars are sourced quadratically, and in the conventional wisdom they should be sourced subdominantly compared to the linear sourcing of GW within perturbation theory. There is copious parameter space, especially for large particle production parameters in which the scalars are even more amplified in the peak of the spectrum compared to GW. This leads to a decrease of the tensor-to-scalar-ratio which is much smaller than 1 at the peak of the signal. As a result $P_h\sim 10^{-2.5}$ needed for PTA result, requires $P_\zeta\gtrsim1$ which violates PBH constraints.
%\end{enumerate}
%

%We further commented on why backreaction constraints allow for larger amplifications for the axion-$U(1)$ model, and 
We show that the axion-$U(1)$ model, for a finite amount of rolling, potentially explains the deviation from astrophysical background in PTA data, and is consistent with the given spectral shape, together with interesting phenomenological signatures such as chiral primordial GW, scalar induced GW and PBH production.
We find that such a parameter region is not excluded but is close to the perturbativity and backreaction bounds derived in small/mild backreaction regime, and thereby there is a need for non-linear analysis.
We also note that there is a clear smoking gun of axion inflation coupled to gauge fields, namely statistical parity violation, hence it is expected that the resultant GW background is almost perfectly chiral \footnote{possibly with a scale dependent chirality \cite{Garcia-Bellido:2023ser}}, which is a concrete prediction for forthcoming surveys
\cite{Crowder:2012ik,Kato:2015bye,Smith:2016jqs,Belgacem:2020nda,Sato-Polito:2021efu}.  
\\
\\
{\bf Acknowledgement.} We are grateful to Robert Caldwell, Angelo Caravano, Emanuela Dimastrogiovanni, Matteo Fasiello, Daniel G. Figueroa and Eiichiro Komatsu for the comments on the draft. C.U. thanks Yann Gouttenoire and Lorenzo Sorbo for the discussions on effective degrees of freedom from gauge fields. C.U. dedicates this work to Harun Kolçak, and thanks his family for their support. C.U. is supported by the Kreitman fellowship of the Ben-Gurion University, and the Excellence fellowship of the Israel Academy of Sciences and Humanities, and the Council for Higher Education. A.P. is supported by IBS under the project code, IBS-R018-D1. I.O. is supported by JSPS KAKENHI Grant No.~JP20H05859 and 19K14702.

\bibliographystyle{apsrev4-1}
\bibliography{Ref.bib}

\end{document}